\documentclass[fleqn,usenatbib]{mnras}

\usepackage{newtxtext,newtxmath}

\usepackage[T1]{fontenc}
\usepackage{ae,aecompl}

\usepackage{graphicx}	
\usepackage{amsmath}	
\usepackage{amssymb}	

\def\f{\frac}
\def\btheta{\boldsymbol{\theta}}

\def\data{\mathbf{d}}
\def\bdtwidle{\mathbf{t}}

\def\mean{\boldsymbol{\mu}}
\def\fisher{\mathbf{F}}
\def\transpose{\mathrm{T}}
\def\L{\mathcal{L}}

\def\transpose{\mathrm{T}}
\def\score{\mathbf{s}}
\def\cov{\mathbf{C}}
\def\tr{\mathrm{tr}}

\title[Generalized massive data compression]{Generalized massive optimal data compression}

\author[J. Alsing, B. Wandelt]{
Justin Alsing$^{1, 2}$\thanks{E-mail: jalsing@flatironinstitute.org}
and Benjamin Wandelt$^{1,3,4,5}$
\\
$^{1}$Center for Computational Astrophysics, Flatiron Institute, 162 5th Ave, New York City, NY 10010, USA\\
$^{2}$Imperial Centre for Inference and Cosmology, Department of Physics, Imperial College London, Blackett Laboratory,\\  Prince Consort
Road, London SW7 2AZ, UK\\ 
$^{3}$Institut d'Astrophysique de Paris (IAP), UMR 7095, CNRS UPMC Universite Paris 6, Sorbonne Universitse,\\  98bis boulevard Arago, F-75014 Paris, France\\
$^{4}$Institut Lagrange de Paris (ILP), Sorbonne Universitse, 98bis boulevard Arago, F-75014 Paris, France\\
$^{5}$Department of Physics and Astronomy, University of Illinois at Urbana-Champaign, 1002 W Green St, Urbana, IL 61801, USA
}

\date{Accepted XXX. Received YYY; in original form ZZZ}

\pubyear{2018}

\begin{document}
\label{firstpage}
\pagerange{\pageref{firstpage}--\pageref{lastpage}}
\maketitle

\begin{abstract}
Data compression has become one of the cornerstones of modern astronomical data analysis, with the vast majority of analyses compressing large raw datasets down to a manageable number of informative summaries. In this paper we provide a general procedure for optimally compressing $N$ data down to $n$ summary statistics, where $n$ is equal to the number of parameters of interest. We show that compression to the score function -- the gradient of the log-likelihood with respect to the parameters -- yields $n$ compressed statistics that are optimal in the sense that they preserve the Fisher information content of the data. Our method generalizes earlier work on linear Karhunen-Lo\'{e}ve compression for Gaussian data whilst recovering both lossless linear compression and quadratic estimation as special cases when they are optimal. We give a unified treatment that also includes the general non-Gaussian case as long as mild regularity conditions are satisfied, producing optimal non-linear summary statistics when appropriate.  As a worked example, we derive explicitly the $n$ optimal compressed statistics for Gaussian data in the general case where both the mean and covariance depend on the parameters.
\end{abstract}




\section{Introduction}
Data analysis problems in astronomy and cosmology generally involve inferring $n$ parameters of interest from $N$ data, where $N$ is typically much larger than $n$. In this paper we are concerned with compressing large datasets down to just $n$ numbers -- one per parameter -- whilst retaining as much information about the parameters as possible. Such massive data compression schemes find widespread application, making subsequent inference from very large datasets tractable. Maximum-likelihood estimation or Bayesian parameter inference can be performed on the likelihood of the compressed statistics with massively reduced computational cost, as shown for linear compression by \cite{Heavens2000a}.

As a new frontier, likelihood-free inference and Approximate Bayesian Computation methods are emerging as a viable approach to analyzing large and complex astronomical datasets \citep{Schafer2012, Cameron2012, Weyant2013, Robin2014, Lin2015, Akeret2015, Ishida2015, Jennings2016, Hahn2017, Kacprzak2017, Carassou2017, Davies2017, Alsing2018a}. These methods generically involve simulating mock data given parameters and comparing the simulated data to the real data, accepting parameters when the mock data is "close" (by some metric) to the real data. This comparison in data-space suffers from the curse of dimensionality, with computational cost scaling exponentially in the size of the data set $N$. Massive data compression is absolutely essential for these methods to be scalable to large datasets.


\citet{Heavens2000a}, following earlier work by \citet{Tegmark1997}, derived an optimal linear data compression scheme for Gaussian data. They derived linear combinations of the data that maximize the Fisher information content of the compressed statistics, finding that in the case where only the mean depends on the parameters the full Fisher information is preserved under linear compression to just $n$ numbers. This radical compression scheme, commonly known as \textsc{moped}, has been applied successfully to a wide range of problems in astronomy and cosmology, including determining star formation histories of galaxies \citep{Reichardt2001, Heavens2004, Panter2007}, cosmic microwave background data analysis \citep{Gupta2002,Zablocki2016}, gravitational waves \citep{Graff2011}, transient detection \citep{Protopapas2005}, fast covariance matrix estimation \citep{Heavens2017}, galaxy power spectrum and bispectrum analyses \citep{Gualdi2017} and more. Beyond linear data compression, optimal quadratic compression has become a standard tool for analyzing cosmological (and other) power spectra \citep{Tegmark1997, Bond1998, Bond2000} with widespread applications; massive optimal data compression has become one of the cornerstones of modern astronomical data analysis. In spite of their successes, linear and quadratic compression are only optimal under very specific circumstances, eg., for Gaussian data and when only the mean or only the covariance depend on the parameters, respectively. Given their widespread use in astronomical data analysis, it is worthwhile generalizing these optimal compression schemes to non-Gaussian likelihood functions with arbitrary parameter dependencies.

In this work we generalize the results of \citet{Tegmark1997} and \citet{Heavens2000a} and describe a general procedure for compressing $N$ data down to $n$ numbers -- one per parameter of interest -- such that the Fisher information is saturated, for any given likelihood function and in a framework that is not restricted to Gaussian data or linear statistics. We also derive explicitly the compressed statistics for Gaussian likelihoods where both the mean and covariance may depend on the parameters, extending the work of \citet{Heavens2000a} to the general case. By phrasing the compression problem in terms of a Taylor expansion of the log-likelihood, we find a simpler and more general derivation of optimal compressed statistics. Our generalized results recover optimal linear compression \citep{Heavens2000a} and optimal quadratic compression \citep{1997PhRvD..55.5895T,Bond1998,Bond2000} as special cases when they are optimal.

The structure of this paper is as follows: in \S \ref{sec:score_compression} we develop a general procedure for compressing $N$ data down to $n$ summary statistics given a likelihood function, without losing information. In \S \ref{sec:gaussian_compression} we derive the compressed statistics for Gaussian likelihood functions where both the mean and covariance depend on the parameters. In \S \ref{sec:llfi} we discuss the utility of the massive compression for likelihood and likelihood-free inference. In \S \ref{sec:failure_modes} we highlight the failure modes of this data compression scheme and discuss how to proceed in these cases. We conclude in \S \ref{sec:conclusions}.
\section{Optimal compression for general likelihood functions}
\label{sec:score_compression}
\subsection{Fisher information and the information inequality}
Our goal is to compress the $N$ data $\data$ down to $n$ numbers $\bdtwidle$, whilst retaining as much information about the parameters as possible. The \emph{information inequality} provides a natural way of defining what we mean by compressed statistics that are "as informative as possible"; for any vector of statistics of the data $\mathbf{t}\in\mathbb{R}^n$, the information inequality gives the lower bound on the variance (see eg., \citealp{Lehmann2006}),
\begin{align}
\label{info_inequality}
\mathrm{Var}_{\boldsymbol\theta}\left[t_\alpha\right] \geq \left(\mathbf{A}^\transpose\,\fisher^{-1}\,\mathbf{A}\right)_{\alpha\alpha},
\end{align}
where\footnote{Here and in the following $\mathrm{E}_{\btheta}[x]$ denotes the expectation of $x$ taken for fixed parameters $\btheta$, and gradients $\nabla$ denote derivatives with respect to $\btheta$.} the $\mathbf{A} = \nabla\,\mathrm{E}_{\btheta}\left[\mathbf{t}^\transpose\right]$  and the \emph{Fisher information matrix} $\fisher$ is defined as 
\begin{align}
\fisher \equiv  - \mathrm{E}_{\boldsymbol\theta}\left[\nabla\nabla^\transpose\L\right] = \mathrm{E}_{\boldsymbol\theta}\left[\nabla\L\nabla^\transpose\L\right].
\end{align}
with the second equality holding under mild regularity conditions (see section \ref{sec:failure_modes}).
Note that the Fisher information is in general a function of the parameters under which the expectation value is taken; in the information inequality Eq. \eqref{info_inequality}, the Fisher information is evaluated at the fiducial parameters $\btheta$.

In the special case where the statistics $\mathbf{t}\in\mathbb{R}^n$ happen to be an unbiased estimator for the parameters, ie. $\mathrm{E}_{\boldsymbol\theta}\left[\mathbf{t}\right]=\btheta$, then the matrices $\mathbf{A}$ become identity matrices and the information inequality reduces to the Cram\'{e}r-Rao bound:
\begin{align}
\label{cramer_rao}
\mathrm{Var}_{\boldsymbol\theta}\left[t_\alpha\right] \geq \fisher^{-1}_{\alpha\alpha}.
\end{align}
In the same spirit as \citet{Tegmark1997} and \citet{Heavens2000a}, we will derive compressed statistics that are "optimal" in the sense that they saturate the lower bound of the information inequality in Eq. \eqref{info_inequality}, evaluated at some fiducial parameter set $\btheta_*$ that we have chosen a priori. In cases where a sensible fiducial point cannot be chosen, it may be necessary to iterate.

\citet{Heavens2000a} derived the $n$ linear combinations of the data that maximize the Fisher information at the fiducial point, leading to an optimal linear data compression of the data. We take a different approach whereby we find sufficient statistics of the linearized log-likelihood function, which we show saturate the lower bound of the information inequality. By phrasing the problem in terms of an expansion of the log-likelihood, we are able to derive a more general prescription for data compression that naturally includes non-linear statistics when necessary, can be applied to any given likelihood function and is readily extended to higher-order sufficient statistics.

The argument proceeds as follows: in \S \ref{sec:sufficient} we derive sufficient statistics for the parameters from the linearized log-likelihood function, and demonstrate that these saturate the information inequality. In \S \ref{sec:quasi} we show that these optimally compressed statistics can be used to form a quasi maximum-likelihood estimator that saturates the Cram\'{e}r-Rao bound and iterates to the true maximum-likelihood estimator, which is important in situations where a satisfactory fiducial parameter set cannot be chosen a priori for performing the data compression.
\subsection{Sufficient statistics of linearized likelihoods and compression to the score function}
\label{sec:sufficient}
Taylor expanding the log-likelihood to second order in the parameters about some fiducial point $\btheta_*$ we have,
\begin{align}
\label{taylor}
\L = \L_* + \delta\btheta^\transpose\nabla\L_* - \f{1}{2}\delta\btheta^\transpose\mathbf{J}_*\delta\btheta,
\end{align}
where the derivative of the log-likelihood is commonly referred to as the \emph{score function} $\score \equiv \nabla\L$, the negative second derivative is the \emph{observed information matrix} $\mathbf{J} \equiv -\nabla\nabla^\transpose\L$, and '$*$' denotes quantities that are evaluated at $\btheta_*$. Both the score function and the observed information are functions of the data.

To linear order in the parameters, the parameters only couple to the data through the score function $\nabla\L_*$ -- a vector of length $n$ and a function of the data. The score function hence constitutes the sufficient statistics for the parameters for the linearized log-likelihood function. This immediately provides a natural data compression from $N$ data down to $n$ compressed numbers, ie., computing only the $n$ data combinations that appear in the score function:
\begin{align}
\label{score}
\bdtwidle = \nabla\L_*.
\end{align}
We can easily show that the score function saturates the lower bound of the information inequality, Eq. \eqref{info_inequality}: Taking the covariance of $\mathbf{t}$ (ie., the left hand side of Eq. \ref{info_inequality}) gives,
\begin{align}
\mathrm{Cov}_{\btheta_*}\left[\mathbf{t},\mathbf{t}\right] = \mathrm{E}_{\btheta_*}\left[\nabla\mathcal{L}_*\nabla^\transpose\mathcal{L}_*\right] = \mathbf{F}_*,
\end{align}
where we have used the fact that $\mathrm{E}_{\btheta_*}\left[\nabla\mathcal{L}_*\right]=0$. Hence, the covariance of $\mathbf{t}$ evaluated for data generated using the fiducial parameters $\btheta_*$ is equal to the Fisher matrix. Now let us consider the right hand side of the information inequality. Using the fact that \begin{align}
\mathbf{A}=\nabla\,\mathrm{E}_{\btheta}\left[\nabla^\transpose\mathcal{L}\right]=\mathrm{E}_{\btheta}\left[\nabla \nabla^\transpose\mathcal{L}\right]=-\mathbf{F},
\label{exchangeDiffAndInt}
\end{align}
the right hand side of the information inequality gives 
$\mathbf{A}_*^\transpose\,\fisher^{-1}_*\,\mathbf{A}_* = \mathbf{F}_*$;
hence, the statistics $\mathbf{t}=\nabla\L_*$ saturate the lower bound of the information inequality Eq. \eqref{info_inequality}, evaluated about the fiducial point $\btheta_*$. Assuming that the assumptions of the information inequality are satisfied (see section \ref{sec:failure_modes}) no other statistics can provide more (Fisher) information.

This strikingly simple result tells us that the score function represents optimally compressed statistics in the sense of saturating the Fisher information. Importantly, the score function will not in general be a linear function of the data; our approach generalizes the linear Karhunen-Lo\'{e}ve compression considered in previous studies \citep{Tegmark1997, Heavens2000a} to give non-linear compressed statistics when appropriate, and is applicable to any likelihood function. 
\subsection{Connection to maximum-likelihood estimation and saturation of the Cram\'{e}r-Rao bound}
\label{sec:quasi}
The data combinations appearing in $\nabla\L_*$ are linearly related to a quasi maximum-likelihood estimator whose covariance is equal to the inverse Fisher information derived from the full likelihood, hence saturating the Cram\'{e}r-Rao bound Eq. \eqref{cramer_rao}. This can be seen as follows: Maximizing the Taylor expanded log-likelihood Eq. \eqref{taylor} with respect to the parameters yields a quasi maximum-likelihood estimator,
\begin{align}
\label{estimator}
\hat\btheta = \btheta_* + \mathbf{J}^{-1}_*\nabla\L_*,
\end{align}
where both the score function $\nabla\L_*$ and the observed information $\mathbf{J}^{-1}_*$ depend on the observed data. In practice, it is useful to replace the observed information matrix with its expectation value, ie., the Fisher information matrix $\fisher_* \equiv \mathrm{E}_{\btheta_*}\left[\mathbf{J}_*\right]$,
\begin{align}
\label{quasi_estimator}
\hat\btheta = \btheta_* + \fisher^{-1}_*\nabla\L_*.
\end{align}
Now the estimator only depends on the data through the score function $\nabla\L_*$, ie., our compressed statistics $\bdtwidle$. The covariance of the quasi maximum-likelihood estimator (evaluated about the expansion point) is given by:
\begin{align}
\mathrm{Cov}_{\btheta_*}\left[\hat\btheta,\hat\btheta\right] &= \fisher^{-1}_*\mathrm{E}_{\btheta_*}\left[\nabla\L_*\nabla^\transpose\L_*\right] \fisher^{-1}_* \nonumber \\
&= \fisher^{-1}_*,
\end{align}
where in the third line we have used the fact that $\mathrm{E}_{\btheta_*}\left[\nabla\L_*\nabla^\transpose\L_*\right] \equiv \fisher_*$. Hence, the covariance of the quasi maximum-likelihood estimator is equal to the Fisher information matrix evaluated at the fiducial parameters and the Cram\'{e}r-Rao bound in Eq. \eqref{cramer_rao} is saturated.

Similarly to the linear compression derived in \citet{Tegmark1997} and \citet{Heavens2000a}, compression to the score function described above only guarantees that the information inequality is saturated about the expansion point. If the expansion point is close to the maximum-likelihood, this procedure will be close to optimal. However, in situations where a suitable expansion point cannot be chosen a priori, it may be necessary to iterate Eq. \eqref{quasi_estimator} towards the maximum-likelihood, ie.,
\begin{align}
\label{iterative_mle}
\hat\btheta_{k+1} = \hat{\btheta}_k + \fisher^{-1}_k\nabla\L_k.
\end{align}
This is the well known \emph{Fisher scoring method} for maximum-likelihood estimation. In the limit $k\rightarrow\infty$ Eq. \eqref{iterative_mle} converges to the maximum-likelihood estimator, which has a number of important properties: it is asymptotically unbiased, and its sampling distribution is asymptotically Gaussian with covariance equal to the Fisher information evaluated at the maximum-likelihood point (rather than at some a priori chosen expansion point as in Eq. \ref{quasi_estimator}). After $k$ iterations, Eq. \eqref{iterative_mle} yields a quasi maximum-likelihood estimator formed exclusively from data combinations constituting $\mathbf{t} = \nabla\L_{k-1}$, that saturates the Fisher information evaluated at the estimated parameters at the previous iteration. 

Note that when the full log-likelihood is quadratic with curvature $\fisher$ in the neighborhood of $\hat\btheta_{k}$ and this neighborhood includes the maximum likelihood estimator, then $\hat\btheta_{k+1}$ will be the maximum likelihood estimator. Since this is approximately the case in many cosmological applications, the first iterate $\hat\btheta_{1}$ is usually an excellent approximation to the maximum likelihood estimator and highly robust to the choice of fiducial parameters.
\subsection*{Summary}
The score function (Eq. \ref{score}) contains  $n$ sufficient statistics of the log-likelihood function expanded to leading order in the $n$ parameters about some fiducial point. It represents  an optimal compression from $N$ data to $n$ numbers, saturating the information inequality. These compressed statistics are linearly related to a quasi maximum-likelihood estimator (Eq. \ref{quasi_estimator}) that saturates the Cram\'{e}r-Rao bound, and can be iterated to the formal maximum-likelihood estimator. Compression to the score or equivalently to the quasi maximum-likelihood estimator hence provides a generic approach to data compression that preserves Fisher information.

In the next section we derive explicitly the compressed score statistics for a Gaussian likelihood function. As we will see, the linear {\sc moped} compression of \citet{Heavens2000a}, and optimal quadratic compression \citep{1997PhRvD..55.5895T,Bond1998,Bond2000}, appear as special cases.
\section{Optimal compression for Gaussian likelihood functions}
\label{sec:gaussian_compression}
In this section we will apply the principles of \S \ref{sec:score_compression} to the case of Gaussian likelihood functions.

Assuming a Gaussian likelihood, we have
\begin{align}
\label{gaussian}
\L = -\f{1}{2}(\data - \mean)^\transpose \cov^{-1}(\data - \mean) - \f{1}{2}\ln|\cov|,
\end{align}
where we will consider the general case where both the mean and covariance depend on the parameters, $\mean = \mean(\btheta)$ and $\cov=\cov(\btheta)$. Taylor expanding the log-likelihood to second order about some fiducial point $\btheta_*$ we have
\begin{align}
\label{gauss_taylor}
\L = \L_* + \delta\btheta^\transpose\nabla\L_* - \f{1}{2}\delta\btheta^\transpose\mathbf{J}_*\delta\btheta,
\end{align}
where the score function is given by
\begin{align}
\label{gauss_score}
\nabla\L = \nabla\mean^\transpose\cov^{-1}(\data - \mean) + \f{1}{2}(\data - \mean)^\transpose \cov^{-1}\nabla \cov\,&\cov^{-1}(\data - \mean) \nonumber \\
&- \f{1}{2}\tr(\cov^{-1}\nabla \cov),
\end{align}
and the observed information matrix is
\begin{align}
\label{gauss_observed_information}
\mathbf{J} = -\nabla\nabla^\transpose\L =& -\nabla\nabla^\transpose\mean^\transpose\cov^{-1}(\data - \mean) 
+ 2\nabla\mean^\transpose\cov^{-1}\nabla^\transpose\cov\,\cov^{-1}(\data - \mean) \nonumber \\
&+ \nabla\mean^\transpose\cov^{-1}\nabla^\transpose\mean
+\nabla\mean^\transpose\cov^{-1}\nabla\cov\,\cov^{-1}(\data - \mean) \nonumber \\
&+(\data - \mean)^\transpose\cov^{-1}\nabla\cov\,\cov^{-1}\nabla^\transpose\cov\,\cov^{-1}(\data - \mean) \nonumber \\
&-\f{1}{2}(\data - \mean)^\transpose\cov^{-1}\nabla\nabla^\transpose\cov\,\cov^{-1}(\data - \mean) \nonumber \\
&+ \f{1}{2}\tr\left(\cov^{-1}\nabla\nabla^\transpose\cov\right)  
 - \f{1}{2}\tr\left(\cov^{-1}\nabla\cov\,\cov^{-1}\cov^{-1}\nabla^\transpose\cov\right),
\end{align}
and we have used the fact that $\nabla\cov^{-1}=\cov^{-1}\nabla\cov\,\cov^{-1}$ and $\nabla\ln|\cov|=\tr(\cov^{-1}\nabla \cov)$.

Taking the expectation of the observed information yields the Fisher information matrix,
\begin{align}
\label{gauss_fisher}
\fisher \equiv -\mathrm{E}\left[\nabla\nabla^\transpose\L\right] = \nabla\mean^\transpose\cov^{-1}\nabla^\transpose\mean + \f{1}{2}\tr\left(\cov^{-1}\nabla\cov\,\cov^{-1}\nabla^\transpose\cov\right),
\end{align}
where we have used the fact that $\mathrm{E}\left[(\data - \mean)\right] = 0$ and $\mathrm{E}\left[(\data - \mean)(\data - \mean)^\transpose\right] = \cov$.

The score function contains the sufficient statistics of the linearized log-likelihood function and defines our compressed statistics. Keeping only the terms that depend on the data, the score Eq. \eqref{gauss_score} gives compressed statistics:
\begin{align}
\label{gauss_dtwidle}
\bdtwidle = \nabla\mean_*^\transpose\cov^{-1}_*(\data - \mean_*) + \f{1}{2}(\data - \mean_*)^\transpose \cov^{-1}_*\nabla \cov_*\,\cov_*^{-1}(\data - \mean_*),
\end{align}
where the covariance of $\mathbf{t}$ is equal to the Fisher information, $\mathrm{Cov}_{\btheta_*}\left[\mathbf{t},\mathbf{t}\right] = \mathbf{F}_*$.

Following \S \ref{sec:score_compression} (Eq. \ref{quasi_estimator}), this allows us to write down a quasi maximum-likelihood estimator for the parameters,
\begin{align}
\label{gauss_quasi}
\hat\btheta = \btheta_* + \fisher_*^{-1}\big[\nabla\mean_*^\transpose\cov_*^{-1}(\data - \mean_*) + \f{1}{2}(\data - \mean_*)&^\transpose \cov_*^{-1}\nabla \cov_*\cov_*^{-1}(\data - \mean_*)\nonumber \\
& - \f{1}{2}\tr(\cov_*^{-1}\nabla \cov_*)\big],
\end{align}
where the data only enters via our compressed statistics, $\bdtwidle$ (Eq. \ref{gauss_dtwidle}). Taking the covariance of the above quasi maximum-likelihood estimator recovers the inverse of the Fisher matrix in Eq. \eqref{gauss_fisher}; compression to either Eq. \ref{gauss_dtwidle} or equivalently the estimator Eq. \ref{gauss_quasi} (or any non-singular linear transformation of it, such as the decorrelated $\fisher^{-1/2}\bdtwidle$) hence saturates the lower bound of the information inequality.

The compressed data vector in Eq. \eqref{gauss_dtwidle} has two terms. The first term is linear in the data and represents a simple linear compression. In the case where only the mean depends on the parameters, ie., $\nabla\cov = 0$, this is the only relevant term and recovers the linear {\sc moped} compression derived in \citet{Heavens2000a}\footnote{Note that \citet{Heavens2000a} derived the linear compression that maximized the Fisher information, under the constraint that the covariance of the compressed statistic is equal to the identity matrix.  This results in the compressed statistics $\fisher^{-1/2}\bdtwidle$ that differ from \eqref{gauss_dtwidle} only through a linear transformation that leaves the Fisher information invariant.}. In this case, the Fisher matrix also reduces to the first term in Eq. \eqref{gauss_fisher}.

The second term is quadratic in the data and represents a quadratic compression. In cases where only the covariance depends on the parameters and the mean is fixed, this will be the only term that survives, and similarly the Fisher matrix reduces to the second term in Eq. \eqref{gauss_fisher}. A classic example where this special case occurs in cosmology when estimating the power spectrum from noisy observations of a Gaussian field (for example, the cosmic microwave background) and gives the optimal quadratic estimator for the power spectrum \citep{1997PhRvD..55.5895T,Bond1998,Bond2000}.

Whilst many situations are covered by either only the mean or only the covariance depending on the parameters, there are many cases where both the mean and covariance are functions of the parameters. In these general cases, both terms in Eqs. \eqref{gauss_dtwidle} and \eqref{gauss_fisher} are required together to give data compression that saturates the information criterion. 

By restricting to linear data compression, \citet{Heavens2000a} derived lossless compression in the case where only the mean depends on the parameters. The more general case where either the covariance or both the mean and covariance depend on the parameters results in a Karhunen-Lo\'{e}ve eigenvalue problem, where $N$ linear combinations of the data  could subsequently be ordered by their eigenvalues so that only the most informative linear combinations are kept. In general this leads to lossy compression \citep{Tegmark1997}. In contrast, our more general approach (with both terms in Eq. \eqref{gauss_dtwidle}) results in compression to $n$ numbers that saturates the information criterion for all cases.
\subsection*{Summary}
In the case of a Gaussian likelihood, the score function defines $n$ compressed statistics of the data, given in Eq. \eqref{gauss_dtwidle}, that constitute our optimal data compression. Equivalently, these compressed statistics can be re-packaged into a quasi maximum-likelihood estimator, given in Eq. \eqref{gauss_quasi}; both contain the same information and saturate the information inequality. The optimal compressed statistics from the Gaussian likelihood contain two terms; one linear and one quadratic in the data. Individually, these terms cover the cases where only the mean and only the covariance depend on the parameters respectively, recovering the {\sc moped} compression and the optimal quadratic estimator. In the general case where both the mean and covariance depend on the parameters, both terms are required to saturate the information inequality.
\section{Likelihood and likelihood-free inference using optimally compressed statistics}
\label{sec:llfi}
The next question is how to do efficient inference once the compressed statistics $\mathbf{t}$ of the data have been calculated. 

For traditional likelihood-based inference, this requires knowing the likelihood function $P(\mathbf{t} | \btheta)$. In the general case, the likelihood for $\mathbf{t}$ is asymptotically Gaussian,
\begin{align}
P(\mathbf{t} | \btheta) = \f{1}{(2\pi)^{\f{n}{2}}|\mathbf{F}|^\f{1}{2}}\mathrm{\exp}\left[-\f{1}{2}(\mathbf{t} - \mean_\mathbf{t}(\btheta))^\transpose\mathbf{F}^{-1}(\mathbf{t} - \mean_\mathbf{t}(\btheta))\right],
\end{align}
where $\mathbf{F}$ is the Fisher matrix, and the mean $\mean_\mathbf{t}(\btheta)$ is the expectation value of the statistics $\mathbf{t}$ for parameters $\btheta$. When compressing under a Gaussian likelihood, the Fisher matrix is given in Eq. \eqref{gauss_fisher} and $\mean_\mathbf{t}(\btheta)$ is given by Eq. \eqref{gauss_dtwidle} but with $\data$ replaced by $\mu(\btheta)$, ie., the data expectation at parameters $\btheta$.

In the general case, or in the case when the covariance of Eq. \eqref{gaussian} depends on the parameters and the compressed statistics includes the term quadratic in the data, the likelihood will only be asymptotically Gaussian. In the special case where only the mean depends on the parameters (corresponding to {\sc moped} compression) the Gaussian likelihood for $\mathbf{t}$ will be exact when $\btheta=\btheta_\ast$. The benefit of performing a likelihood-based analysis with the compressed statistics rather than the full dataset is clear; for large datasets, the cost of individual likelihood evaluations can be reduced enormously, often by many orders of magnitude. A classic example is compression of CMB maps down to a quadratic estimator for the power spectrum; the likelihood function over the full $n_\mathrm{pix}\sim 10^6$ pixel map requires inverting an $n_\mathrm{pix}\times n_\mathrm{pix}$ covariance matrix, whilst the likelihood for the few thousand optimally compressed power spectrum modes is asymptotically Gaussian and many orders of magnitude cheaper to compute (once the summaries have been calculated; \citealp{Tegmark1997}). For high-resolution maps the former is computationally intractable in practice, whilst the latter is possible thanks to the massive data compression. The generalized results of this paper will allow for massive optimal compression under unrestricted likelihood assumptions.

For application to likelihood-free inference, the likelihood for the compressed statistics is not needed; the only requirement is that one can forward simulate realizations of $\mathbf{t}$. Likelihood-free inference is typically applied in situations where the true likelihood function is not available; in these cases, compression can be performed under an approximate likelihood and the resulting compressed statistics will only be optimal to the extent that this is a good approximation to the true likelihood. If a ``best guess'' for the true likelihood is made for the compression, then the compression can be thought of as optimal up to our state of knowledge about the true likelihood; you can only do as well as your likelihood-ignorance allows. 

In practice, raw data will often initially be compressed to some heuristic summaries that are thought to contain the physical information; for example, cosmological surveys are typically compressed down to estimated power spectra, with the number of estimated power spectrum modes still $\gg n$. If these ``first-level summaries" are informative about the parameters, then when subsequently compressing them down to just $n$-numbers, one might reach for a Gaussian likelihood approximation with the justification that the first-level summaries will be asymptotically Gaussian under the central limit theorem. As a new frontier, optimal compression without likelihood approximations may be achieved by training information maximizing neural networks on forward simulations \citep{Charnock2018}.

Typically, the score-compression requires knowledge of some statistical properties of the data; for example, compression under a Gaussian likelihood in Eq. \eqref{gauss_dtwidle} requires knowledge of the data mean, covariance and their derivatives at some fiducial point. When these are not known a priori, they can be estimated from forward simulations, or if an approximate model for the data means and covariances is available this can be used. Crucially, any approximations made in the compression step can only result in sub-optimality and will not bias the subsequent likelihood-free inference. This means compression to the score provides a general and robust compression scheme for likelihood-free inference. For a detailed discussion of implementing score-compression for likelihood-free inference, see \citet{Alsing2018a}.
\section{Limitations, failure modes and extensions}
\label{sec:failure_modes}
There are special cases where compression from $N$ data down to $n$ numbers can catastrophically fail, leading to incorrect parameter inferences. As pointed out by \citet{Graff2011}, whenever the compression leads to a many-to-one mapping from $\data\rightarrow\bdtwidle$ one can get spurious additional modes in the posterior inference from the compressed data. This makes good intuitive sense: if there are two parameter sets that lead to identical predictions for the compressed statistics, this will lead to two peaks in the compressed likelihood surface, even if one of those parameter sets would have been strongly disfavored by the original likelihood. In these many-to-one compression cases, one can break the degeneracies and build a one-to-one compression by including two or more sets of $n$ compressed statistics computed about different expansion points $\btheta_*$ \citep{Protopapas2005}. 

The information inequality holds under weak conditions, namely that the score function is defined for all $d$ in the support of the likelihood and that the order of expectation and differentiation can be swapped in equation \eqref{exchangeDiffAndInt}. These mild assumptions also underly our results. 

The Fisher information measures the expected curvature of the likelihood. In cases where we are far from the asymptotic limit where the maximum-likelihood estimator is Gaussian with covariance given by the Fisher matrix, the Fisher information is not guaranteed to be a good measure of the information content of the likelihood. In these cases, the approach developed in \S \ref{sec:score_compression} provides a natural framework for extending our data compression scheme to include additional informative statistics: for example one can  continue the Taylor expansion of Eq. \eqref{taylor} and find sufficient statistics of the log-likelihood expanded to higher orders in the parameters.
\section{Conclusions}
\label{sec:conclusions}
The score function -- the derivative of the log-likelihood with respect to the parameters, $\mathbf{t}=\nabla\L$ -- constitutes sufficient statistics for the log-likelihood expanded up to linear order in the parameters. This provides a natural way of compressing $N$ data down to $n$ summary statistics (one per parameter); furthermore, we have shown that compression to the score is optimal in the sense that it saturates the information inequality. This provides a general procedure for optimally compressing data down to $n$ summaries, for a given likelihood. Our results generalize earlier work on separate optimal linear and quadratic compression from Gaussian likelihood functions.

By phrasing the problem in terms of sufficient statistics of the log-likelihood function expanded to a given order in the parameters, we found both a simple route to the optimal compression, and also a clear path to extending the compression beyond saturation of the Fisher information: continuing the expansion to higher orders in the parameters.
\section*{Acknowledgements}
We thank Stephen Feeney for useful discussions. Benjamin Wandelt is supported by the Labex ILP (reference ANR-10-LABX-63). This work is supported by the Simons Foundation.
\bibliographystyle{mnras}
\bibliography{data_compression}

\begin{thebibliography}{}
\makeatletter
\relax
\def\mn@urlcharsother{\let\do\@makeother \do\$\do\&\do\#\do\^\do\_\do\%\do\~}
\def\mn@doi{\begingroup\mn@urlcharsother \@ifnextchar [ {\mn@doi@}
  {\mn@doi@[]}}
\def\mn@doi@[#1]#2{\def\@tempa{#1}\ifx\@tempa\@empty \href
  {http://dx.doi.org/#2} {doi:#2}\else \href {http://dx.doi.org/#2} {#1}\fi
  \endgroup}
\def\mn@eprint#1#2{\mn@eprint@#1:#2::\@nil}
\def\mn@eprint@arXiv#1{\href {http://arxiv.org/abs/#1} {{\tt arXiv:#1}}}
\def\mn@eprint@dblp#1{\href {http://dblp.uni-trier.de/rec/bibtex/#1.xml}
  {dblp:#1}}
\def\mn@eprint@#1:#2:#3:#4\@nil{\def\@tempa {#1}\def\@tempb {#2}\def\@tempc
  {#3}\ifx \@tempc \@empty \let \@tempc \@tempb \let \@tempb \@tempa \fi \ifx
  \@tempb \@empty \def\@tempb {arXiv}\fi \@ifundefined
  {mn@eprint@\@tempb}{\@tempb:\@tempc}{\expandafter \expandafter \csname
  mn@eprint@\@tempb\endcsname \expandafter{\@tempc}}}

\bibitem[\protect\citeauthoryear{{Akeret}, {Refregier}, {Amara}, {Seehars}  \&
  {Hasner}}{{Akeret} et~al.}{2015}]{Akeret2015}
{Akeret} J.,  {Refregier} A.,  {Amara} A.,  {Seehars} S.,   {Hasner} C.,  2015,
  \jcap, 8, 043

\bibitem[\protect\citeauthoryear{Alsing, Wandelt  \& Feeney}{Alsing
  et~al.}{2018}]{Alsing2018a}
Alsing J.,  Wandelt B.,   Feeney S.,  2018, arXiv preprint arXiv:1801.01497

\bibitem[\protect\citeauthoryear{Bond, Jaffe  \& Knox}{Bond
  et~al.}{1998}]{Bond1998}
Bond J.,  Jaffe A.~H.,   Knox L.,  1998, Physical Review D, 57, 2117

\bibitem[\protect\citeauthoryear{Bond, Jaffe  \& Knox}{Bond
  et~al.}{2000}]{Bond2000}
Bond J.,  Jaffe A.~H.,   Knox L.,  2000, The Astrophysical Journal, 533, 19

\bibitem[\protect\citeauthoryear{Cameron \& Pettitt}{Cameron \&
  Pettitt}{2012}]{Cameron2012}
Cameron E.,  Pettitt A.,  2012, Monthly Notices of the Royal Astronomical
  Society, 425, 44

\bibitem[\protect\citeauthoryear{Carassou, de Lapparent, Bertin  \&
  Borgne}{Carassou et~al.}{2017}]{Carassou2017}
Carassou S.,  de Lapparent V.,  Bertin E.,   Borgne D.~L.,  2017, arXiv
  preprint arXiv:1704.05559

\bibitem[\protect\citeauthoryear{Charnock, Lavaux  \& Wandelt}{Charnock
  et~al.}{2018}]{Charnock2018}
Charnock T.,  Lavaux G.,   Wandelt B.,  2018, arXiv preprint arXiv:1802.03537

\bibitem[\protect\citeauthoryear{Davies, Hennawi, Eilers  \& Luki{\'c}}{Davies
  et~al.}{2017}]{Davies2017}
Davies F.~B.,  Hennawi J.~F.,  Eilers A.-C.,   Luki{\'c} Z.,  2017, arXiv
  preprint arXiv:1703.10174

\bibitem[\protect\citeauthoryear{Graff, Hobson  \& Lasenby}{Graff
  et~al.}{2011}]{Graff2011}
Graff P.,  Hobson M.~P.,   Lasenby A.,  2011, Monthly Notices of the Royal
  Astronomical Society: Letters, 413, L66

\bibitem[\protect\citeauthoryear{Gualdi, Manera, Joachimi  \& Lahav}{Gualdi
  et~al.}{2017}]{Gualdi2017}
Gualdi D.,  Manera M.,  Joachimi B.,   Lahav O.,  2017, arXiv preprint
  arXiv:1709.03600

\bibitem[\protect\citeauthoryear{Gupta \& Heavens}{Gupta \&
  Heavens}{2002}]{Gupta2002}
Gupta S.,  Heavens A.~F.,  2002, Monthly Notices of the Royal Astronomical
  Society, 334, 167

\bibitem[\protect\citeauthoryear{Hahn, Vakili, Walsh, Hearin, Hogg  \&
  Campbell}{Hahn et~al.}{2017}]{Hahn2017}
Hahn C.,  Vakili M.,  Walsh K.,  Hearin A.~P.,  Hogg D.~W.,   Campbell D.,
  2017, Monthly Notices of the Royal Astronomical Society, 469, 2791

\bibitem[\protect\citeauthoryear{Heavens, Jimenez  \& Lahav}{Heavens
  et~al.}{2000}]{Heavens2000a}
Heavens A.~F.,  Jimenez R.,   Lahav O.,  2000, Monthly Notices of the Royal
  Astronomical Society, 317, 965

\bibitem[\protect\citeauthoryear{Heavens, Panter, Jimenez  \& Dunlop}{Heavens
  et~al.}{2004}]{Heavens2004}
Heavens A.,  Panter B.,  Jimenez R.,   Dunlop J.,  2004, Nature, 428, 625

\bibitem[\protect\citeauthoryear{Heavens, Sellentin, de Mijolla  \&
  Vianello}{Heavens et~al.}{2017}]{Heavens2017}
Heavens A.~F.,  Sellentin E.,  de Mijolla D.,   Vianello A.,  2017, Monthly
  Notices of the Royal Astronomical Society, 472, 4244

\bibitem[\protect\citeauthoryear{Ishida et~al.,}{Ishida
  et~al.}{2015}]{Ishida2015}
Ishida E.,  et~al., 2015, Astronomy and Computing, 13, 1

\bibitem[\protect\citeauthoryear{Jennings, Wolf  \& Sako}{Jennings
  et~al.}{2016}]{Jennings2016}
Jennings E.,  Wolf R.,   Sako M.,  2016, arXiv preprint arXiv:1611.03087

\bibitem[\protect\citeauthoryear{Kacprzak, Herbel, Amara  \&
  R{\'e}fr{\'e}gier}{Kacprzak et~al.}{2017}]{Kacprzak2017}
Kacprzak T.,  Herbel J.,  Amara A.,   R{\'e}fr{\'e}gier A.,  2017, arXiv
  preprint arXiv:1707.07498

\bibitem[\protect\citeauthoryear{Lehmann \& Casella}{Lehmann \&
  Casella}{2006}]{Lehmann2006}
Lehmann E.~L.,  Casella G.,  2006, Theory of point estimation.
Springer Science \& Business Media

\bibitem[\protect\citeauthoryear{Lin \& Kilbinger}{Lin \&
  Kilbinger}{2015}]{Lin2015}
Lin C.-A.,  Kilbinger M.,  2015, Astronomy \& Astrophysics, 583, A70

\bibitem[\protect\citeauthoryear{Panter, Jimenez, Heavens  \& Charlot}{Panter
  et~al.}{2007}]{Panter2007}
Panter B.,  Jimenez R.,  Heavens A.~F.,   Charlot S.,  2007, Monthly Notices of
  the Royal Astronomical Society, 378, 1550

\bibitem[\protect\citeauthoryear{{Protopapas}, {Jimenez}  \&
  {Alcock}}{{Protopapas} et~al.}{2005}]{Protopapas2005}
{Protopapas} P.,  {Jimenez} R.,   {Alcock} C.,  2005, \mnras, 362, 460

\bibitem[\protect\citeauthoryear{Reichardt, Jimenez  \& Heavens}{Reichardt
  et~al.}{2001}]{Reichardt2001}
Reichardt C.,  Jimenez R.,   Heavens A.~F.,  2001, Monthly Notices of the Royal
  Astronomical Society, 327, 849

\bibitem[\protect\citeauthoryear{Robin, Reyl{\'e}, Fliri, Czekaj, Robert  \&
  Martins}{Robin et~al.}{2014}]{Robin2014}
Robin A.,  Reyl{\'e} C.,  Fliri J.,  Czekaj M.,  Robert C.,   Martins A.,
  2014, Astronomy \& Astrophysics, 569, A13

\bibitem[\protect\citeauthoryear{Schafer \& Freeman}{Schafer \&
  Freeman}{2012}]{Schafer2012}
Schafer C.~M.,  Freeman P.~E.,  2012, in , Statistical Challenges in Modern
  Astronomy V.
Springer, pp 3--19

\bibitem[\protect\citeauthoryear{{Tegmark}}{{Tegmark}}{1997}]{1997PhRvD..55.5895T}
{Tegmark} M.,  1997, \prd, 55, 5895

\bibitem[\protect\citeauthoryear{Tegmark, Taylor  \& Heavens}{Tegmark
  et~al.}{1997}]{Tegmark1997}
Tegmark M.,  Taylor A.~N.,   Heavens A.~F.,  1997, The Astrophysical Journal,
  480, 22

\bibitem[\protect\citeauthoryear{Weyant, Schafer  \& Wood-Vasey}{Weyant
  et~al.}{2013}]{Weyant2013}
Weyant A.,  Schafer C.,   Wood-Vasey W.~M.,  2013, The Astrophysical Journal,
  764, 116

\bibitem[\protect\citeauthoryear{Zablocki \& Dodelson}{Zablocki \&
  Dodelson}{2016}]{Zablocki2016}
Zablocki A.,  Dodelson S.,  2016, Physical Review D, 93, 083525

\makeatother
\end{thebibliography}

\bsp	
\label{lastpage}
\end{document}